\documentclass[11pt]{article}
\usepackage{moriond,epsfig}

\def\be{\begin{equation}}
\def\ee{\end{equation}}
\def\bea{\begin{eqnarray}}
\def\eea{\end{eqnarray}}

\begin{document}
\title{REALISTIC OPERATION OF AN ELECTRON ENTANGLER : A DENSITY MATRIX APPROACH}

\author{\underline{OLIVIER SAURET}$^1$, DENIS 
FEINBERG$^1$, THIERRY MARTIN$^2$}

\address{$^1$ LEPES, CNRS, BP 166, 38042 Grenoble, FRANCE}
\address{$^2$ CPT, CNRS, Univ. M\'editerran\'ee, Case 907, 13288 
Marseille, FRANCE}

\maketitle\abstracts{
 The detailed operation of an electron spin entangler is studied, using 
density matrix equations. 
The device is made of a superconductor, two quantum dots and two normal 
leads. The treatment takes 
into account coherent tunneling in a non-perturbative way, and analyzes 
the various parasitic effects,
in addition to the main process (crossed Andreev reflection) : those 
include singlet pairs passing 
through a single dot, or cotunneling between dots through the 
superconductor. The optimum 
operation of the device is characterized.}

Producing entangled electron pairs is a challenge for fundamental 
experiments 
(analogous to those performed with photons\cite{photons}), as well 
as controlling quantum information\cite{IQ} in solid state 
devices\cite{loss_divincenzo}. 
It has been proposed that a superconductor, connected to energy filters, 
can 
serve as a source of spin singlets\cite{lesovik_martin_blatter,loss}. Here 
we address the 
operation of the "S-DD entangler" made of a superconductor (S) connected 
in parallel 
to a double quantum dot (DD) 
where Coulomb blockade prevents two electrons to pass through a single 
dot. The dots are small enough so 
that a single electron state is involved in each dot. The "crossed 
Andreev" (CA)
\cite{crossedAndreev,choi_bruder_loss} process is thus favoured, where a 
spin singlet is emitted, 
shared by the DD. The feasibility of such a device crucially depends on 
the control 
of "parasitic" processes spoiling entanglement, mainly of two kinds : 
first, singlet pairs can pass through 
a single dot, either through a double occupation state (direct Andreev 
DA), or one by one (Fig. 1). Second,
an electron can pass from one dot to the other by elastic 
cotunneling\cite{crossedAndreev,CT} (CT) through S. 
The first, but not the second, was considered in Ref. 5. In 
addition, all 
processes are mixed together, making a consistent treatment difficult. 
Such a study is indeed possible by using the density matrix equations, which 
generalize the usual 
master equations to the inclusion of coherent processes. Those correspond 
to both Andreev transitions
 or to cotunneling. They are made of 
two virtual transitions, with a quasiparticle created in S then destroyed. 
On the contrary,
single electron transitions between dots and the leads L,R are incoherent. 
The complete quantum master (QM) equations for the subsystem made of S and 
the two dots can
be derived\cite{nous} for instance following Ref. 10. 
Here the discussion is based
on the analysis of the dot populations and averaged current flow. Further 
results are devoted to 
shot noise correlations and Bell inequalities\cite{inprep}. Notice that 
the S-DD entangler was 
recently studied\cite{buttiker} 
in series with a splitter detecting entanglement\cite{burkard}. On the 
other hand, 
QM equations were employed\cite{saraga} for a 
different principle of entangler using another dot instead of a 
superconductor\cite{oliver}. Also, QM equations were at the basis of the analysis of 
a device permitting teleportation of the electron spin in a dot array\cite{TP}. 

In the ideal operation of the S-DD entangler,
the Coulomb blockade in each dot is strong enough so as to rule out 
double occupancy.  
Starting from an empty state $00$, CA reflection allows 
transitions to the singlet state, shared between the two dots, $11_s$ 
with a rate $\gamma_A T$ ($\gamma_A$ is the geometrical factor 
\cite{crossedAndreev,choi_bruder_loss}). 
For a resonant CA process, the dot energy levels satify $\varepsilon=E_1+E_2=0$. 
The two electrons are evacuated from the dots into the reservoirs 
(with chemical potentials $\mu_{L,R}<E_1,E_2$) and the transitions to states $01$, $10$
occur with rates $\Gamma_i$ ($i=L,R$). 

If the Coulomb charging energy is not so strong, a coherent transition 
from $00$ to a doubly 
occupied dot states $20$ or $02$ can occur via a direct Andreev (DA) 
process, which has a 
rate $T_i$ which is larger than for the CA process (Fig. 1). 
Electrons can subsequently 
be detected 
into this reservoir, with rates $\Gamma'_i$. This 
conduction channel 
implies dot energies $E_i+U_i$ associated with double occupancy. One may 
also 
start from an initial state $10$ or $01$. DA can then proceed through 
the empty  
dot, but the charging energy of states $21$ or $12$ equals $E_1+E_2+ U_i$. 
Detection 
in the reservoir can either lead to $20$ ($02$), or to singlet and triplet 
states $11_{s,t}$. Another parasitic channel involves two electrons of a 
Cooper pair tunneling
one by one towards the same reservoir (Fig. 1). This involves a singly occupied 
virtual state which costs 
an energy $\Delta_S$, the superconducting gap. Contrary to the DA process, the dot is emptied 
before the 
quasiparticle in S is anihilated.
This process happens with a rate 
$\tilde{\Gamma}_{i}=\Gamma_i T_i^2/\Delta_S^2$.
Last but not least, cotunneling (CT) allows a coherent transfer of an 
electron from one 
dot to the other 
via S. It couples states $01$ and $10$, but also $20$ ($02$) and $11$,
$21$ and $12$. CT has a rate $\gamma_C T$ which is reduced by a 
geometrical factor.

\begin{figure}[thpb]
\centerline{\epsfig{file=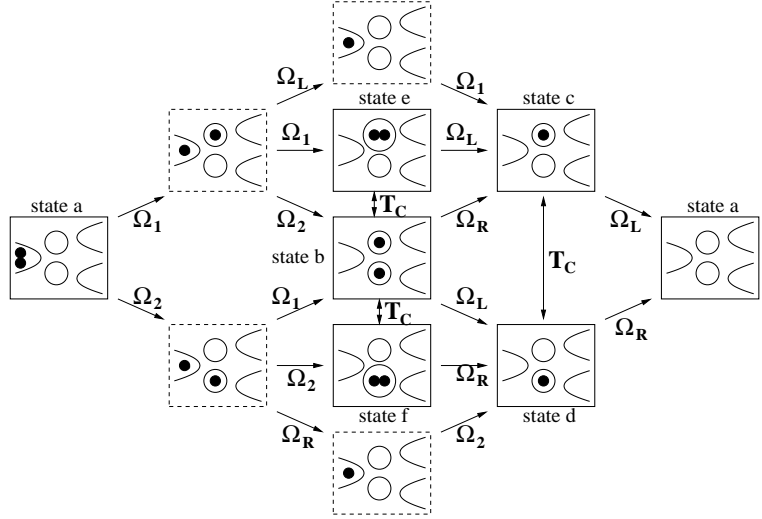,width=10cm,angle=0}} 
\caption{Full operation of the entangler, including the three Andreev 
channels and cotunneling. 
Real states are squared while virtual states are dashed squared. States 
with three 
electrons are omitted for clarity, spin is not represented. The $\Omega$'s 
are the 
one-electron tunneling matrix elements. The crossed Andreev (CA) process is in the middle, 
the direct Andreev ones (DA) second from top or bottom, and the one-by-one processes are 
the top or bottom ones. Cotunneling (CT) connects the states vertically.} 
\label{total_recall}
\end{figure}

Let us exclude high energy states $22$ with $N=4$ electrons in the 
double dot (DD), 
keeping states $00$, $\sigma0$, $0\sigma$ (with spin $\sigma$), $11$ 
(singlet and triplets), 
$2\sigma$ and $\sigma2$ ($2$ means a local singlet). Transport 
of electron pairs through the DD is highly correlated. With CA alone, 
pairs pass 
one after the other through the DD. Including the parasitic processes, 
there is a strong mixing 
of processes and no simple perturbative calculation is possible. Yet, 
starting from the full 
one-electron Hamiltonian, one can derive a complete 
set of QM equations for the populations $p_{\alpha}=\rho_{\alpha\alpha}$ 
and the 
"coherences" $\rho_{\alpha\beta}$ 
(diagonal and non-diagonal matrix elements of the density matrix 
$\rho_{DD}$). It takes the general form \cite{gurvitz} :

\begin{eqnarray}
\dot\rho_{\alpha\alpha} &=&
i\sum_{\beta}T_{\alpha\beta}(\rho_{\alpha\beta}-\rho_{\beta\alpha})
- \sum_{\gamma}(\Gamma_{\alpha\gamma}\rho_{\alpha\alpha}
- \Gamma_{\gamma\alpha}\rho_{\gamma\gamma})\\
\dot\rho_{\alpha\beta} &=& i(E_{\alpha}-E_{\beta})\rho_{\alpha\beta} + 
i\sum_{\gamma}(\rho_{\alpha\gamma}
T_{\beta\gamma} -\rho_{\gamma
\beta}T_{\alpha \gamma}) -\frac{\rho_{\alpha\beta}}{2}
\sum_{\gamma} (\Gamma_{\alpha \gamma} +  \Gamma_{\beta \gamma})
\end{eqnarray}

\noindent
where the $\Omega$'s are the coherent rates and the $\Gamma$'s the 
incoherent rates. 
One assumes in the derivation that virtual states with at most one 
quasiparticle in S are created. 
The obtained set of equations\cite{nous} is valid up to second order in 
the matrix 
elements describing tunneling to the leads, and to any order in the 
coherent rates 
(which are of order two in the tunneling matrix elements between the 
superconductor and the dots). 

In Ref. 5, a T-matrix calculation was performed, calculating 
separately the ideal (CA) 
current, and the DA and one-by-one parasitic currents. Here the optimum 
operation of the device can be 
settled on a firmer basis, and a better understanding of the physics 
involved is obtained. 
First let us assume a symmetric device ($\Gamma_L = \Gamma_R$) and treat 
the processes separately, 
without cotunneling. Assuming  $\gamma_A^2T^2\gg\varepsilon^2$, the CA 
current in each lead is

\begin{equation} \label{current_ent_approx}
I_L^{CA}\approx 
e\Gamma\frac{8\gamma_A^2T^2}{8\gamma_A^2T^2+\Gamma^2/4}
\end{equation}

On the other hand, the DA current $I^{DA}$ and the one-by-one current 
$I^{obo}$ read

\begin{equation}
I_L^{DA}=e\Gamma\frac{16T^2}{16T^2+\Gamma^2+U^2};\;\;\;\;
I_L^{obo}=e\frac{KT^2\Gamma}{\Delta_S^2+KT^2}
\end{equation}

where $K$ is a numerical constant. If $T << U, \Delta_S$, one has 

\begin{equation}
I_L^{DA}\approx e\Gamma\frac{16T^2}{U^2};\;\;\;\;
I_L^{obo}\approx
4e\Gamma\frac{T^2}{\pi^2\Delta_S^2}
\end{equation}

The general case can be treated, setting $\varepsilon=E_1+E_2$ and putting all processes together. 
As an illustration, 
an analytical formula can be given for the total current, up to first order in the parasitic processes 

\begin{equation}\label{courant}
\begin{array}{rcl}
\nonumber
I_L=e\sigma_0[
\Gamma_L+\Gamma_R+4\Gamma^2\left(\frac{1}{\Gamma_L}-\frac{1}{\Gamma_R}\right) \frac{\gamma_CT^2}{E^2}
-2KA\sigma_0\Gamma_L\frac{T^2}{\Delta_S^2}\left(1-\frac{1}{2\sigma_0}\right)\\
-8A\sigma_0\Gamma_L\frac{T^2}{U^2}\left(5-\frac{1}{\sigma_0}\right)
-2\sigma_0\Gamma_L\frac{\gamma_C^2T^2}{U^2}\left(5-\frac{1}{\sigma_0}+
\frac{2\Gamma\Gamma_L}{\Gamma^2+8\gamma_A^2T_A^2+\varepsilon^2}\right)]
\end{array}
\end{equation}

where $A=\frac{8\gamma_A^2T_A^2}{\Gamma^2+8\gamma_A^2T_A^2+\varepsilon^2}$ 
and
$\sigma_0^{-1}=A+1+\Gamma_R/\Gamma_L+\Gamma_L/\Gamma_R$.

\begin{figure}[thpb]
\centerline{\epsfig{file=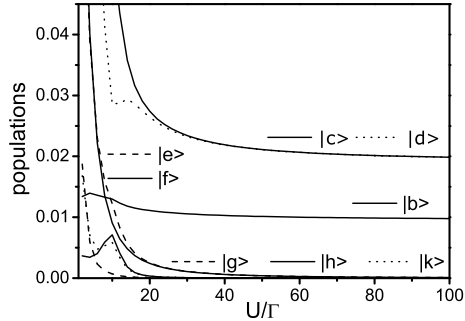,width=7cm,angle=0}}
\caption{Populations of the DD states as a function of $U/\Gamma$ for 
$\Delta_S=9.5K, E_1=-E_2=0.5K, 
\Gamma=T=0.1K, \gamma_A, \gamma_C=0.2$. States b, c, d, e, f, g, h, k 
respectively correspond to charge states 
($11_s$),($10$),($01$),($20$),($02$),($21$),($12$) and ($11_t$).}
\label{TP}
\end{figure} 

The optimization of the entangler requires that most pairs emitted by the 
superconductor leave separated into $L$ and $R$. To fight DA and one-by-one processes, one 
may compare the currents, taken individually for each process, or alternatively adopt a dynamical argument : 
starting from state $|a\rangle=$ ($00$) at time $t=0$, 
the probability $\rho_{11}$ of singlet state $|b\rangle$ oscillates slowly (with frequency 
$\gamma_AT$), but with a large amplitude. On the contrary, the probability of state $|e\rangle=$ ($20$) 
oscillates more rapidly (with frequency 
$T$), but with a small amplitude. The competition between the two processes crucially depends on the decay rate 
$\Gamma$. If it is small, CA is favoured, but if it is too large,  DA process wins, state $|b\rangle$ 
has no time to form. A detailed analysis gives the criterion 
$U,\Delta_S >> max[T,\Gamma/\gamma_A,\varepsilon/\gamma_A]$. A similar analysis can be made for the effect 
of cotunneling : once in the state ($10$) or ($01$), decay in $L,R$ must be faster than the cotunneling 
frequency $\gamma_CT$, leading to the criterion $\gamma_CT<<max[|E_1-E_2|,\Gamma]$. This can be confirmed by
a numerical solution for the probabilities of various states (Fig. 2). 

In summary, density matrix (quantum master) equations can be derived from a microscopic Hamiltonian for a realistic 
entangler, and allow to integrate all processes in a coherent 
and non-perturbative way. A range of parameters for optimum operation is 
$\gamma_AT,\gamma_CT << \Gamma_{L,R} << U,\Delta_S$. More information can be obtained by further analysis 
of the current fluctuations (shot noise correlations)\cite{inprep}.


\section*{References}
\begin{thebibliography}{99}
\bibitem{photons} A. Aspect, J. Dalibard, and G. Roger,
 Phys.\ Rev.\ Lett.\ {\bf 49}, 1804 (1982); L. Mandel, Rev.  Mod.
Phys. {\bf71}, S274 (1999);
A. Zeilinger, Rev.  Mod. Phys. {\bf71}, S288 (1999).
\bibitem{IQ} D. Bouwmeester, A. Ekert, and A.
Zeilinger, {\it The Physics of Quantum Information} (Springer-Verlag,
Berlin, 2000); M. A. Chang and I. L. Nielsen, {\it Quantum Computation and 
Quantum Information}, 
Cambridge University Press, 2000).
\bibitem{loss_divincenzo} D. Loss and D. P. DiVincenzo
Phys. Rev. A {\bf 57}, 120 (1998).
\bibitem{lesovik_martin_blatter} G. B. Lesovik, T. Martin, and G. Blatter, 
Eur.\ Phys.\ J.\ B {24}, 287 (2001); N. M. Chtchelkatchev, G. Blatter, G. B. Lesovik 
and T. Martin, Phys. Rev. B {\bf 66}, 161320 (2002).
\bibitem{loss} P. Recher P., E. V. Sukhorukov, and D. Loss,
Phys. Rev. B {\bf 63}, 165314 (2001).
\bibitem{crossedAndreev} G. Falci, D. Feinberg, and F. W. J. Hekking,
Europhys.  Lett. {\bf 54}, 255 (2001)
\bibitem{choi_bruder_loss} M. S. Choi, C. Bruder, and D. Loss,
 Phys. Rev. B{\bf 62}, 13569 (2000).
\bibitem{CT} D. V. Averin and Yu.  V. Nazarov, in {\it Single Charge 
Tunneling}, H. Grabert and M.H. Devoret eds.
(Plenum, New York 1992). G. Bignon, M. Houzet, F. Pistolesi and F. 
Hekking, cond-mat/0310349 (2003).
\bibitem{nous} O. Sauret, D. Feinberg and T. Martin, cond-mat/0402416 
(Phys. Rev. B, 
to appear). 
\bibitem{gurvitz} S. A. Gurvitz and Ya.  S. Prager, Phys.
Rev. B {\bf 53}, 15932 (1996).
\bibitem{inprep} O. Sauret, T. Martin and D. Feinberg, in preparation.
\bibitem{buttiker} P. Samuelsson, E.V. Sukhorukov and M. B\"uttiker 
cond-mat/0402368.
\bibitem{burkard} G. Burkard, D. Loss, and E. V. Sukhorukov Phys. \ Rev. \ 
B {\bf61}, R16303 (2000). 
\bibitem{saraga} D. Saraga, D. Loss, Phys. Rev. Lett. 90, 
166803 (2003). 
\bibitem{oliver}  W. D. Oliver, F. Yamaguchi, and Y. Yamamoto, Phys. Rev. 
Lett. {\bf 88}, 037901 (2002).
\bibitem{TP} O. Sauret, D. Feinberg and T. Martin, Eur. Phys. J. B {\bf 
32}, 545 (2003); ibid., 
Phys. Rev. B 69,035332 (2004). 
\end{thebibliography}
\end{document}